\documentclass[sigconf]{acmart}

\usepackage{graphicx}
\usepackage{balance}  

\usepackage{caption}
\usepackage{subcaption}
\usepackage{soul}
\usepackage[linesnumbered,ruled]{algorithm2e}
\usepackage{algorithmic}
\usepackage{ifthen}
\usepackage{amsfonts}

\usepackage{booktabs} 
\usepackage{enumitem}
\usepackage{comment}

\usepackage{hyperref}

\usepackage{multirow}

\theoremstyle{definition}
\newtheorem{definition}{Definition}[section]

\usepackage{relsize}

\usepackage{slashbox}

\usepackage{setspace}
\usepackage{microtype}

\usepackage{xcolor}

\setcopyright{rightsretained}

\usepackage{xcolor,colortbl}

\definecolor{Gray}{gray}{0.70}
\definecolor{Gray2}{gray}{0.90}
\definecolor{LightCyan}{rgb}{0.88,1,1}

\newcolumntype{b}{>{\columncolor{Gray}}c}
\newcolumntype{a}{>{\columncolor{Gray2}}c}
\newcolumntype{d}{>{\columncolor{LightCyan}}c}

\acmConference[UrbComp 2019]{International Workshop on Urban Computing }{August 2019}{Anchorage, Alaska - USA}
\acmYear{2019}
\copyrightyear{2019}

\acmArticle{4}
\acmPrice{15.00}

\begin{document}
\title{A Countrywide Traffic Accident Dataset}
\titlenote{All rights reserved to the authors, and The Ohio State University (2019).}

\author[ ]{Sobhan Moosavi}
\affiliation{%
  \institution{The Ohio State University}
  \institution{Department of Computer Science and Engineering}
  \streetaddress{2015 Neil Ave}
  \city{Columbus}
  \state{Ohio}
  \postcode{43210}
}
\email{moosavi.3@osu.edu}

\author[ ]{Mohammad Hossein Samavatian}
\affiliation{%
  \institution{The Ohio State University}
  \institution{Department of Computer Science and Engineering}
  \streetaddress{2015 Neil Ave}
  \city{Columbus}
  \state{Ohio}
  \postcode{43210}
}
\email{samavatian.1@osu.edu}

\author[ ]{Srinivasan Parthasarathy}
\affiliation{%
  \institution{The Ohio State University}
  \institution{Department of Computer Science and Engineering}
  \streetaddress{2015 Neil Ave}
  \city{Columbus}
  \state{Ohio}
  \postcode{43210}
}
\email{srini@cse.ohio-state.edu}

\author[ ]{Rajiv Ramnath}
\affiliation{%
  \institution{The Ohio State University}
  \institution{Department of Computer Science and Engineering}
  \streetaddress{2015 Neil Ave}
  \city{Columbus}
  \state{Ohio}
  \postcode{43210}
}
\email{ramnath@cse.ohio-state.edu}

\begin{abstract}
Reducing traffic accidents is an important public safety challenge. However, the majority of studies on traffic accident analysis and prediction have used small-scale datasets with limited coverage, which limits their impact and applicability; and existing large-scale datasets are either private, old, or do not include important contextual information such as environmental stimuli (weather, points-of-interest, etc.). In order to help the research community address these shortcomings we have - through a comprehensive process of data collection, integration, and augmentation - created a large-scale publicly available database of accident information named \textit{US-Accidents}. US-Accidents currently contains data about $2.25$ million instances of traffic accidents that took place within the contiguous United States, and over the last three years. Each accident record consists of a variety of intrinsic and contextual attributes such as {\em location}, {\em time}, {\em natural language description}, {\em weather}, {\em period-of-day}, and {\em points-of-interest}. We present this dataset in this paper, along with a wide range of insights gleaned from this dataset with respect to the spatiotemporal characteristics of accidents. The dataset is publicly available at \url{https://smoosavi.org/datasets/us_accidents}.
\end{abstract}


\maketitle

\section{Introduction}
\label{sec:intro}
Reducing traffic accidents is an important public safety challenge around the world. A global status report on traffic safety \cite{world2015global}, notes that there were 1.25 million traffic deaths in 2013 alone, with deaths increasing in 68 countries when compared to 2010. Accident prediction is important for optimizing public transportation, enabling safer routes, and cost-effectively improving the transportation infrastructure, all in order to make the roads safer. Given its significance, accident analysis and prediction has been a topic of much research in the past few decades. While a large body of research has been focused on small-scaled datasets with limited coverage (e.g. a small number of road-segments, or just one city) \cite{chang2005analysis,caliendo2007crash,lin2015novel,wenqi2017model}, the value and impact of predictive solutions may be better studied when using large-scale data. Although some studies conducted their work based on large-scale motor-vehicle crash datasets, their data is usually private or poses strict rules to be shared with outside researchers, which makes their framework and results unproducible \cite{eisenberg2004mixed,yuan2017predicting,chen2016learning,najjar2017combining}. While there are still a few publicly available large-scale accident datasets, their data is either old, limited to one state or one city, or incomprehensive (regarding data attributes or average reports per year) \cite{uk_accidents,ny_accidents,ia_accidents,md_accidents,seattle_accidents}.

In order to mitigate these challenges and to provide a context for future research on traffic accident analysis and prediction, we present a new dataset, we name it {\em US-Accidents}, which includes about $2.25$ million instances of traffic accidents took place within the contiguous United States\footnote{The contiguous United States excludes Alaska and Hawaii, and considers District of Columbia (DC) as a separate state.} between February 2016 and March 2019. Unlike some of the available large-scale accident datasets (such as \cite{ny_accidents}), US-Accidents offers a wide range of data attributes to describe each accident record including {\em location data}, {\em time data}, {\em natural language description of event}, {\em weather data}, {\em period-of-day information}, and {\em relevant points-of-interest data} (traffic signal, stop sign, etc.). Very importantly, we also present our  \textit{process} for creating the above dataset from streaming traffic reports and heterogeneous contextual data (weather, points-of-interests, etc.), so that the community can validate it, and with the belief that this process can itself serve as a model for dataset creation. 

Using US-Accidents, we performed a variety of data analysis and profiling to derive a wide-range of insights. Our analyses demonstrated that 
about $40\%$ of accidents took place on or near high-speed roadways (highways, interstates, etc.) and about $32\%$ on or near local roads (streets, avenues, etc.). We also derived various insights with respect to the correlation of accidents with time, points-of-interest, and weather conditions. 

We summarize the contributions of this paper as follows: 
\begin{itemize}[leftmargin=5pt]
    \item A unique dataset, made publicly available at \url{https://smoosavi.org/datasets/us\_accidents}. This dataset has been collected for the contiguous United States over three years, and contains about $2.25$ million traffic accident records. Further, the raw accident records have been augmented by {\em map-matching}, and contextual information such as {\em weather}, {\em period-of-day}, and {\em points-of-interest}. 
    \item A new methodology for the heterogeneous data collection, cleansing, and augmentation; needed to prepare a unique large-scale dataset of traffic accidents. 
    \item A variety of insights gleaned through analyses of accident hot-spot locations, time, weather, and points-of-interest correlations with the accident data; that may directly be utilized for applications such as urban planning, exploring flaws in transportation infrastructure design, traffic management and prediction, and personalized insurance.  
\end{itemize}

The rest of the paper is organized as follows. Section~\ref{sec:rel} provides an overview of related work, followed by definitions and preliminaries in Section~\ref{sec:prob}. The process of creating the accident dataset is presented in Section~\ref{sec:data_process}, and analyses and insights are discussed in Section~\ref{sec:us_accidents}. Finally, Section~\ref{sec:conclusion} concludes the paper. 
\section{Related Work}
\label{sec:rel}
Accident analysis and prediction has an active research topic during the past few decades, with a large body of research has been focused on using small-scale datasets with limited coverage of a few road-segments or one city \cite{chang2005analysis,chang2005data,caliendo2007crash,abellan2013analysis,lin2015novel,kumar2015data,wenqi2017model}. Chang et al. \cite{chang2005analysis} used information such as road geometry, annual average daily traffic, and weather data to predict frequency of accidents for a highway road using a neural network model. Kumar et al. \cite{kumar2015data} applied data mining techniques to extract association rules to perform causality analysis using a small-scale dataset. Likewise, Wenqi et al. \cite{wenqi2017model} applied a convolutional neural network model to perform accident prediction on a road-segment. Although the insights and findings look interesting, the employed datasets are of limited scale; hence, the applicability and generalizability of results might be questionable. 

There are, of course, numerous studies that have used larger-scale datasets  \cite{eisenberg2004mixed,yuan2017predicting,yuan2018hetero,chen2016learning,tamerius2016precipitation,najjar2017combining}; however, the datasets have been either private or not easily accessible. Eisenberg \cite{eisenberg2004mixed} conducted a thorough analysis on the impact of precipitation on road accidents, using a large-scale dataset of about 456,000 crashes collected from 1975 to 2000 for 48 states of the US. Recent studies by Yuan et al. \cite{yuan2018hetero} and Najjar et al. \cite{najjar2017combining} have also employed very large-scale accident datasets to perform real-time traffic accident prediction. However, in neither study have details been shared regarding how the data used may be obtained by others in order to reproduce results for wider use.

Finally, and speaking of datasets alone, there are several publicly available motor vehicle crash datasets; however, they suffer from the limited coverage (e.g., one city or one state) \cite{baton_rouge,ia_accidents,ny_accidents,md_accidents,seattle_accidents}, or their data is old \cite{uk_accidents}, or the provided attributes are not comprehensive enough (missing location, time, or weather data) \cite{ny_accidents,ia_accidents}. To address these challenge, we propose a new process to collect and build a new large-scale accident dataset, with countrywide coverage, and comprehensive data attributes including location, time, weather, period-of-day, and points-of-interest annotations (e.g., intersections, junctions, and traffic signals). 
\section{Terminology}
\label{sec:prob}
In this section we provide a set of definitions.  

\begin{definition}[Traffic Event]
    We define a traffic event $e$ by $e = \langle lat, lng, time, type, desc \rangle$, where $lat$ and $lng$ are GPS latitude and longitude, $type$ is the type of the event, and $desc$ provides a natural language description of the event. A traffic event is of one of the following types: {\em accident}, {\em broken-vehicle}\footnote{Refers to the situation when there is one (or more) disabled vehicle(s) in a road.}, {\em congestion}\footnote{Refers to the situation when the speed of traffic is slower than the expected speed.}, {\em construction}\footnote{An on-going construction or maintenance project on a road.}, {\em event}\footnote{Situations such as {\em sports event}, {\em concerts}, or {\em demonstrations}, that could potentially impact traffic flow.}, {\em lane-blocked}\footnote{Refers to the cases when we have blocked lane(s) due to traffic or weather condition.}, or {\em flow-incident}\footnote{Refers to all other types of traffic events. Examples are {\em broken traffic light} and {\em animal in the road}.}.
\end{definition}

\begin{definition}[Weather Observation Record]
    A weather observation $w$ is defined by $w = \langle lat, lng, time, temperature,$ $humidity,$ $pressure, visibility, wind-speed, precip,$ $rain, snow, fog,$ $hail \rangle$. Here $lat$ and $lng$ represent the GPS coordinates of the weather station which reported $w$; {\em precip} is the precipitation amount (if any); and rain, snow, fog, and hail\footnote{The case of having solid precipitation including ice pallets and hail.} are binary indicators of these events. 
\end{definition}

\begin{definition}[Point-of-Interest]
    A point-of-interest $p$ is defined by $p = \langle lat, lng, type \rangle$. Here, $lat$ and $lng$ show the GPS latitude and longitude coordinates, and available types for $p$ are described in Table~\ref{tab:poi_types}. Note that several of definitions in this table are adopted from \url{https://wiki.openstreetmap.org}. 
\end{definition}

\begin{table}[ht]
    \small
    \setlength\tabcolsep{1pt}
    \centering
    \caption{Definition of Point-Of-Interest (POI) annotation tags based on Open Street Map (OSM).}
    \begin{tabular}{| c | c|}
        \rowcolor{Gray}
        \hline
        \textbf{Type} &  \textbf{Description}\\
        \hline
        Amenity & \begin{tabular}{@{}c@{}} Refers to particular places such as restaurant,\\ library, college, bar, etc.\end{tabular} \\ 
        \hline
        Bump & Refers to speed bump or hump to reduce the speed. \\ 
        \hline
        Crossing & \begin{tabular}{@{}c@{}} Refers to any crossing across roads for \\ pedestrians, cyclists, etc.\end{tabular} \\ 
        \hline
        Give-way & A sign on road which shows priority of passing. \\ 
        \hline
        Junction & Refers to any highway ramp, exit, or entrance. \\ 
        \hline
        No-exit & \begin{tabular}{@{}c@{}} Indicates there is no possibility to travel further\\ by any transport mode along a formal path or route.\end{tabular}  \\ 
        \hline
        Railway & Indicates the presence of railways. \\ 
        \hline
        Roundabout & Refers to a circular road junction.\\  
        \hline
        Station & Refers to public transportation station (bus, metro, etc.). \\ 
        \hline
        Stop & Refers to stop sign. \\ 
        \hline
        Traffic Calming & Refers to any means for slowing down traffic speed. \\ 
        \hline
        Traffic Signal & Refers to traffic signal on intersections. \\ 
        \hline
        Turning Loop & \begin{tabular}{@{}c@{}} Indicates a widened area of a highway with\\a non-traversable island for turning around. \end{tabular}  \\ 
        \hline
    \end{tabular}
    \label{tab:poi_types}
\end{table}

\section{Dataset Creation Process}
\label{sec:data_process}
\begin{figure}[ht]
    \centering
    \includegraphics[scale=0.44]{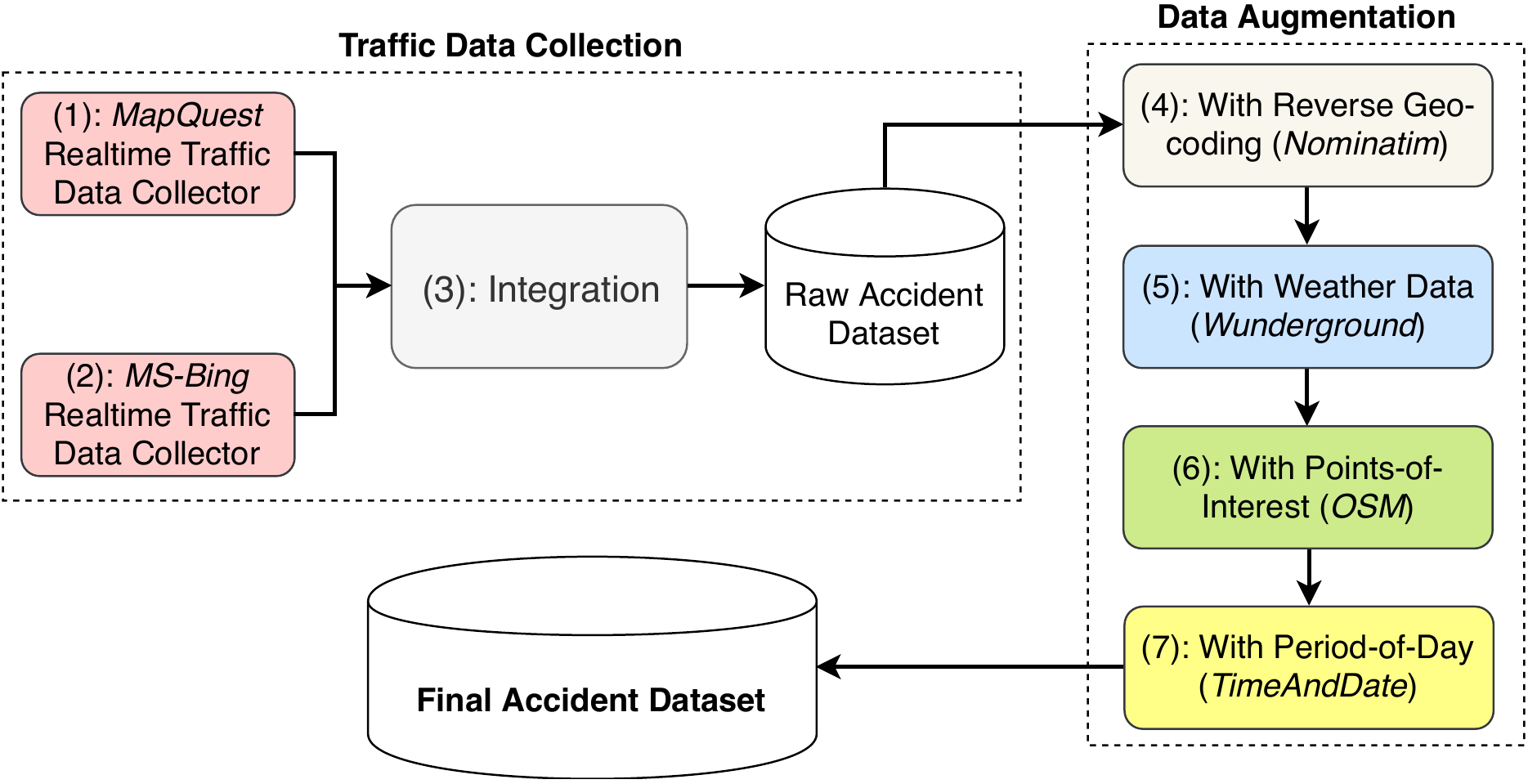}
    \caption{Process of Creating Traffic Accident Dataset}
    \label{fig:data_process}
\end{figure}

An overview of the dataset creation process is shown in Figure~\ref{fig:data_process}, with the following sub-sections provide detailed descriptions of each step. 

\subsection{Traffic Data Collection}

\subsubsection{Realtime Traffic Data Collection}
We collected streaming traffic data using two real-time data providers, namely ``MapQuest Traffic'' \cite{mapquest} and ``Microsoft Bing Map Traffic'' \cite{bing}, whose APIs broadcast traffic events (accident, congestion, etc.) captured by a variety of entities - the US and state departments of transportation, law enforcement agencies, traffic cameras, and traffic sensors within the road-networks. We pulled data every 90 seconds from 6am to 11pm, and every 150 seconds from 11pm to 6am. In total, we collected $2.27$ million cases of traffic accidents between February 2016 and March 2019; $1.7$ million cases were pulled from MapQuest, and $0.54$ million cases from Bing. 

\subsubsection{Integration}
Integration of the data consisted of removing cases duplicated across the two sources and building a unified dataset. We considered two events as duplicates if their Haversine distance and their recorded times of occurrence were both below a heuristic threshold (set empirically at 250 meters and 10 minutes, respectively). We believe these settings to be conservative, but we settled on them in order to ensure a very low possibility of duplicates. Using these settings, we found about $24,600$ duplicated accident records, or about $1\%$ of all data. The final dataset after removing the duplicated cases comprised $2.25$ million accidents. 

\subsection{Data Augmentation}
\subsubsection{Augmenting with Reverse Geo-Coding}
Raw traffic accident records contain only GPS data. We employed the {\em Nominatim} tool \cite{nominatim} to perform reverse geocoding to translate GPS coordinates to addresses, each consisting of a {\em street number}, {\em street name}, {\em relative side (left/right)}, {\em city}, {\em county}, {\em state}, {\em country}, and {\em zip-code}. This process is same as {\em point-wise map-matching}. 

\subsubsection{Augmenting with Weather Data}
Weather information provides important context for traffic accidents. Thus, we employed {\em Weather Underground} API  \cite{wunderground} to obtain  weather information for each accident. Raw weather data was collected from 1,977 weather stations located in airports all around the United States. The raw data comes in the form of observation records, where each record consists of several attributes such as {\em temperature}, {\em humidity}, {\em wind speed}, {\em pressure}, {\em precipitation} (in millimeters), and {\em condition}\footnote{Possible values are {\em clear}, {\em snow}, {\em rain}, {\em fog}, {\em hail}, and {\em thunderstorm}.}. For each weather station, we collected several data records per day, each of which was reported upon any significant change in any of the measured weather attributes. 

Each traffic event $e$ was augmented with weather data as follows. First the closest weather station $s$ was identified. Then, of the weather observation records which are reported from $s$, we looked for the weather observation record $w$ whose reported time was closest to the start time of $e$, and augmented it with weather data. In our integrated accident dataset, the average difference in report time for an accident record and its paired weather observation record was about $15$ minutes; and the $50^{th}$, $75^{th}$, and $90^{th}$ percentiles on time difference distribution were about $11$, $20$, and $26$ minutes, respectively. 

\subsubsection{Augmenting with Points-Of-Interest}
Points-of-interest (POI) are locations annotated on a map as {\em amenities}, {\em traffic signals}, {\em crossings}, etc. These annotations are associated with {\em nodes} on a road-network. A node can be associated with a variety of POI types, however, in this work we only use a subset of 13 types described in Table~\ref{tab:poi_types}. We obtain these annotations from Open Street Map (OSM) \cite{osm} for the United States, using its most recently released dataset (extracted on April 2019). 
The applicable POI annotations for a traffic accident $a$ are based on the actual POI located within a distance threshold $\tau$ from $a$. We determine this threshold by evaluating different threshold values to find the value that is best able to associate a POI with an accident. Essentially, the objective is to find the best distance for which a POI annotation can be identified as a relevant to an accident record. Therefore, we need a mechanism to measure the relevancy. 
To begin with, we note that the natural language descriptions of traffic accidents follow a set of regular expression patterns, and that a few of these patterns may be used to identify and use as an annotation for the location type (e.g., intersection or junction) of the accident. 

\vspace{3pt}
\noindent \textbf{Regular Expression Patterns.} Given the description of traffic events of type accident, we were able to identify 27 regular expression patterns; 16 of them were extracted based on MapQuest data, and 11 from Bing data. Among the MapQuest patterns, the following expression corresponds to {\em junctions} (see Table~\ref{tab:poi_types}): ``$\dots$ \textbf{on} $\dots$ \textbf{at exit} $\dots$'', and the following pattern mostly\footnote{Regarding the distribution of data and using 200 random cases which were manually checked on a map, about 78\% of matches using this pattern were actually happened on intersections.} determines an {\em intersection}: ``$\dots$ \textbf{on} $\dots$ \textbf{at} $\dots$''. We consider a location an \textit{intersection} if it is associated with at-least one of the following annotations (see Table~\ref{tab:poi_types}): {\em crossing}, {\em stop}, or {\em traffic signal}. Among Bing regular expression patterns, two of them identify junctions: ``\textbf{at} $\dots$ \textbf{exit} $\dots$'' and ``\textbf{ramp to} $\dots$''. Table~\ref{tab:accident_examples} shows several examples of accidents, where the regular expression pattern (in bold face) identifies the correct POI type\footnote{These cases were also manually checked on a map to ensure the correctness of the annotation.}. 

\begin{table}[t]
    \small
    \centering
    \setlength\tabcolsep{2pt}
    \caption{Examples of traffic accidents with their {\em annotation type} assigned using their natural language description by regular expression patterns.}
    \begin{tabular}{ c | c | c }
        \textbf{Source} & \textbf{Description} & \textbf{Type}\\
        \hline
        MapQuest & Serious accident \textbf{on} 4th Ave \textbf{at} McCullaugh Rd. & Intersection\\
        MapQuest & Accident \textbf{on} NE-370 Gruenther Rd \textbf{at} 216th St. & Intersection\\
        MapQuest & Accident \textbf{on} I-80 \textbf{at Exit} 4A Treasure Is. & Junction\\
        MapQuest & Accident \textbf{on} I-87 I-287 Southbound \textbf{at Exit} 9 I-287. & Junction\\
        Bing & \textbf{At} Porter Ave/\textbf{Exit} 9 - Accident. Left lane blocked. & Junction\\
        Bing & \textbf{At} IL-43/Harlem Ave/\textbf{Exit} 21B - Accident. & Junction\\
        Bing & \textbf{Ramp to} I-15/Ontario Fwy/Cherry Ave - Accident. & Junction\\
        Bing & \textbf{Ramp to} Q St - Accident. Right lane blocked. & Junction\\
    \end{tabular}
    \label{tab:accident_examples}
\end{table}

\begin{algorithm}[ht]
    \caption{Find Annotation Correlation}
    \begin{algorithmic}[1]
        \STATE Input: a dataset of traffic accidents $\mathcal{A}$, a database of points-of-interest $\mathcal{P}$, and a distance threshold $\tau$.
        \STATE Extract and create a set of regular expression patterns $RE$ to identify a specific POI $\nu$. 
        \STATE Create set $S_1$: for each traffic accident $a \in \mathcal{A}$, we add it to $S_1$ if its natural language description $a.desc$ can be matched with at least one regular expression in set $RE$. 
        \STATE Create set $S_2$: for each traffic accident $a \in \mathcal{A}$, we add it to $S_2$ if there is at least one POI $p \in \mathbf{P}$ of type $\nu$, where $\textit{haversine\_distance }(a, p) \leq \tau$.
        \STATE Output: Return $\textit{Jaccard }(S_1, S_2)$. 
    \end{algorithmic}
    \label{alg:find_tau}
\end{algorithm}

\begin{figure*}[ht]
    \centering
    \hspace{-25pt}
    \begin{subfigure}[b]{0.3\textwidth}
            \includegraphics[width=\linewidth]{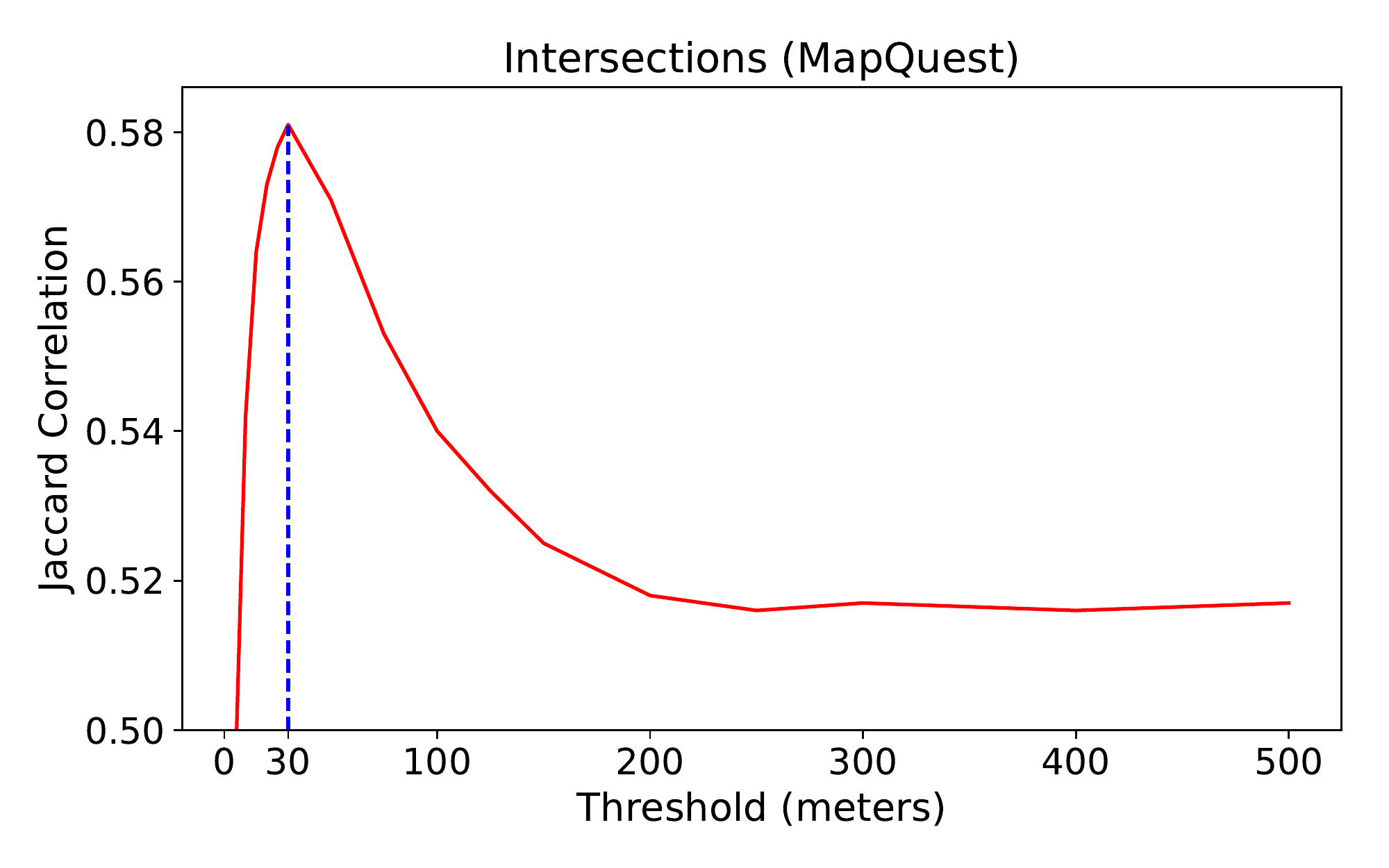}
            \caption{Using MapQuest for Intersection}
            \label{fig:mq_intersec}
    \end{subfigure}\hspace{-5pt}
    \begin{subfigure}[b]{0.3\textwidth}
            \includegraphics[width=\linewidth]{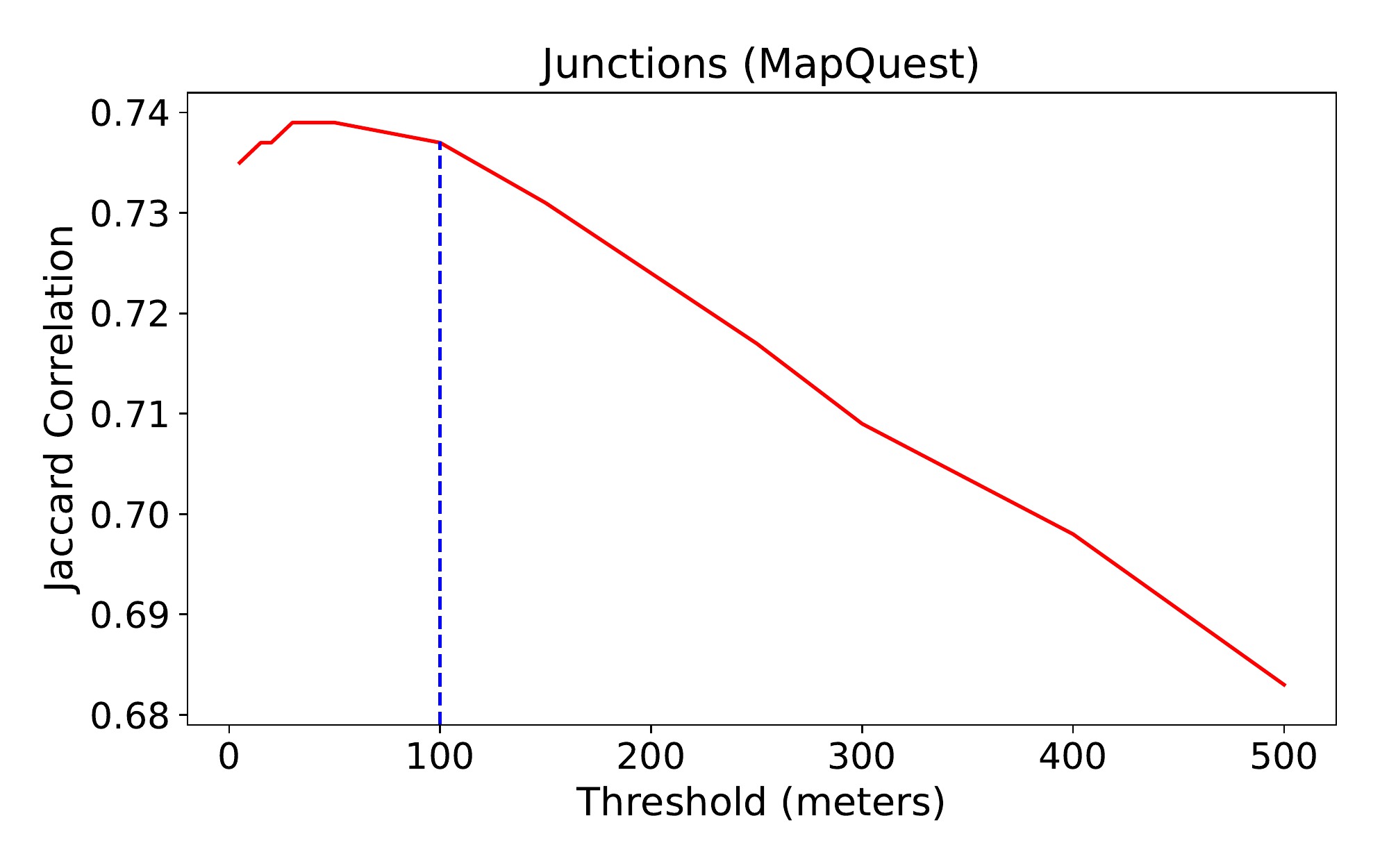}
            \caption{Using MapQuest for Junction}
            \label{fig:mq_junction}
    \end{subfigure}\hspace{-5pt}
    \begin{subfigure}[b]{0.3\textwidth}
            \includegraphics[width=\linewidth]{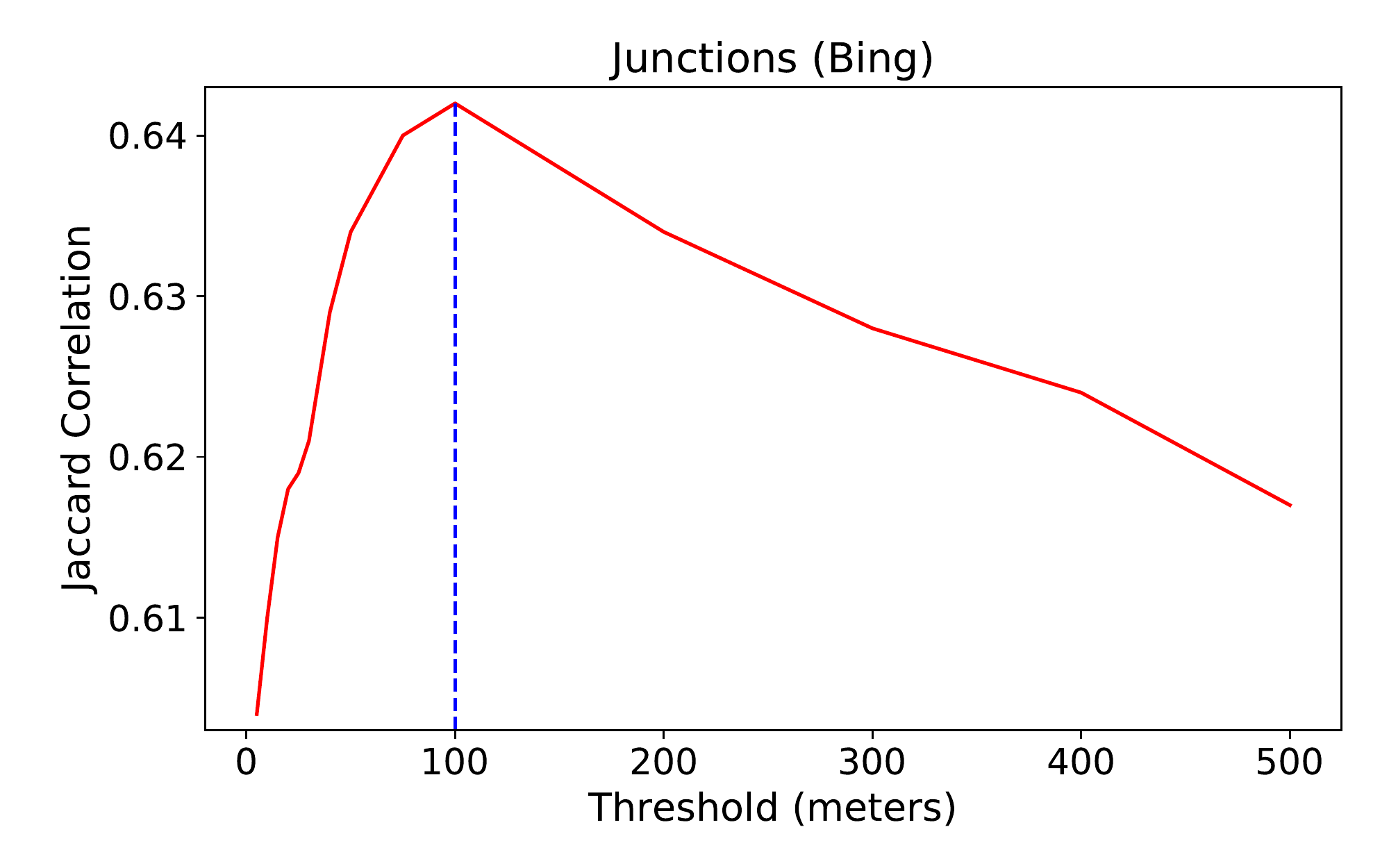}
            \caption{Using Bing for Junction}
            \label{fig:bg_junction}
    \end{subfigure}
    \hspace{-22pt}
    \caption{Correlation study between regular-expression and OSM-based extracted annotations to find the best distance threshold values.}
    \label{fig:text_tag_correlation}
\end{figure*}

The  essential idea is to find a threshold value that maximizes the correlation between annotations from POI and annotations derived using regular expression patterns. Thus, for a set of accident records, we annotate their location based on both regular expression patterns as well as OSM-based POI annotations (using a specific distance threshold). Then, we measure the correlation between the annotations derived from these methods to find which threshold value provides the highest correlation (i.e., the best choice). Note that we employ the regular expression patterns as \textit{pseudo} ground truth labels, to evaluate OSM-based POI annotations using different threshold values. We propose Algorithm~\ref{alg:find_tau} to find the best distance threshold. We use a sample of $100,000$ accidents as set $\mathcal{A}$ (step 1). For step 2, we consider either ``intersection'' or ``junction'', and use the set of relevant regular expressions (see Table~\ref{tab:accident_examples}) in terms of $RE$. Next we create set $S_1$ by annotating each traffic accident $a \in \mathcal{A}$ by using the regular expression patterns in $RE$ (step 3). Then we annotate each traffic accident $a \in \mathcal{A}$ based on points-of-interests in $\mathcal{P}$, using the distance threshold $\tau$ to create $S_2$ (step 4). Finally, we calculate the Jaccard similarity score using Equation~\ref{eq:jaccard} (step 5):
\begin{equation}
    \label{eq:jaccard}
    \textit{Jaccard }(S_1, S_2) = \frac{\big|\, S_1 \cap S_2\, \big|}{\big|\, S_1 \cup S_2\, \big|}
\end{equation}
We examined the following candidate set to find the optimal threshold value (all values in meters): $\{5, 10, 15, 20, 25, 30, 40, 50, 75, 100, 125,$ $150, 200, 250, 300, 400, 500\}$. We separately studied samples from Bing and MapQuest, and employed corresponding regular expression patterns for ``intersection'' and ``jucntion''. Figure~\ref{fig:text_tag_correlation} shows the results for each data source and each annotation type. From Figure~\ref{fig:mq_intersec}, we see that the maximum correlation for intersections is obtained for a threshold value of 30 meters. Figures~\ref{fig:mq_junction} and \ref{fig:bg_junction} show that 100 meters is an appropriate distance threshold for annotating a junction. 

Thresholds for the other types of available annotations in Table~\ref{tab:poi_types} are derived from the thresholds for junction and intersection as described below:
\begin{itemize}[leftmargin=*]
    \item \textbf{Junction-based threshold.} Given the definition of a junction (i.e., a highway ramp, exit, or entrance), we used the same threshold for the following types: amenity and no-exit. 
    \item \textbf{Intersection-based threshold.} Given the definition of an intersection, we used the same threshold for the following annotation types: bump, crossing, give-way, railway, roundabout, station, stop, traffic calming, traffic signal, and turning loop. 
\end{itemize}
Using these thresholds, we augmented each accident record with points-of-interest. In summary, $27.5 \%$ of accident records were augmented with at least one of the available POI types in Table~\ref{tab:poi_types}. Further discussion on annotation results are presented in Section~\ref{sec:us_accidents}.  

\subsubsection{Augmenting with Period-of-Day}
Given the start time of an accident record, we used ``TimeAndDate'' API \cite{timeanddate} to label it as {\em day} or {\em night}. We assign this label based on four different daylight systems, namely {\em Sunrise/Sunset}, {\em Civil Twilight}, {\em Nautical Twilight}, and {\em Astronomical Twilight}. Note that these systems are defined based on the position of the sun with respect to the horizon\footnote{See \url{https://en.wikipedia.org/wiki/Twilight} for more details.}. 

\section{US-Accidents Dataset}
\label{sec:us_accidents}
\begin{figure*}
    \minipage{0.32\textwidth}
        \centering
        \includegraphics[width=\linewidth]{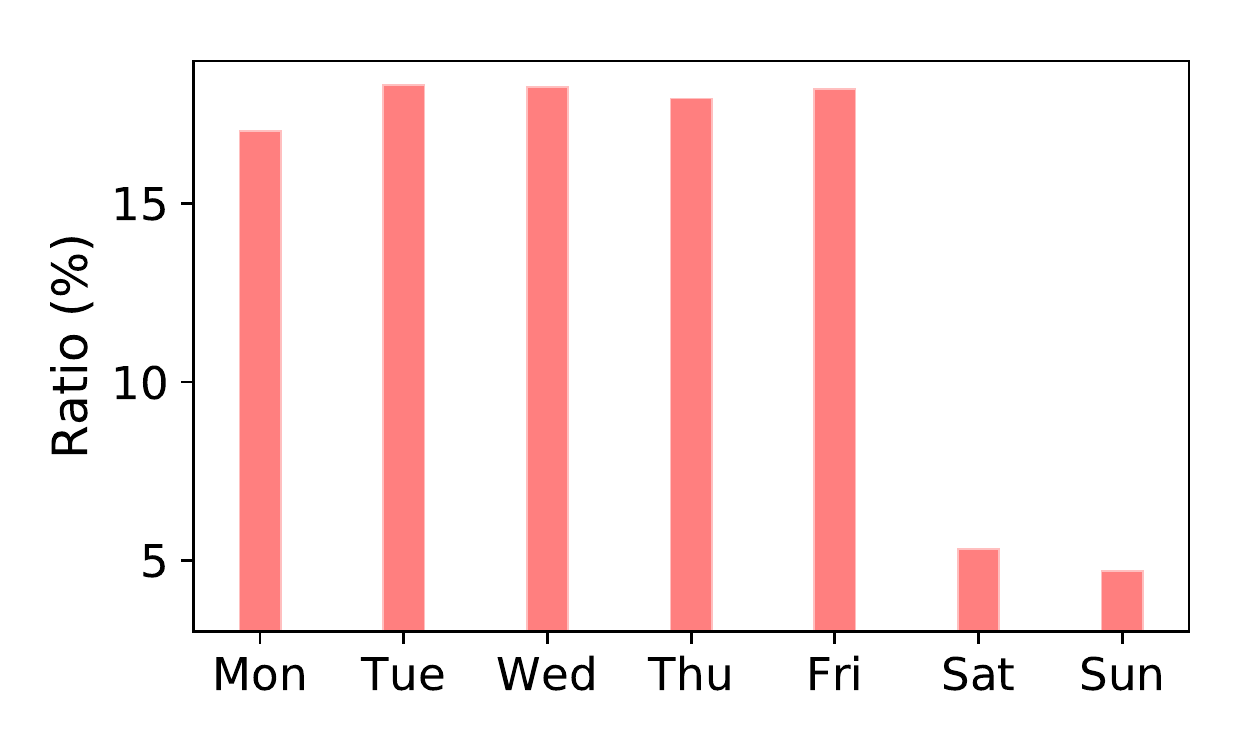}
        (a) Day of Week
    \endminipage\hfill
    \minipage{0.32\textwidth}
        \centering
        \includegraphics[width=\linewidth]{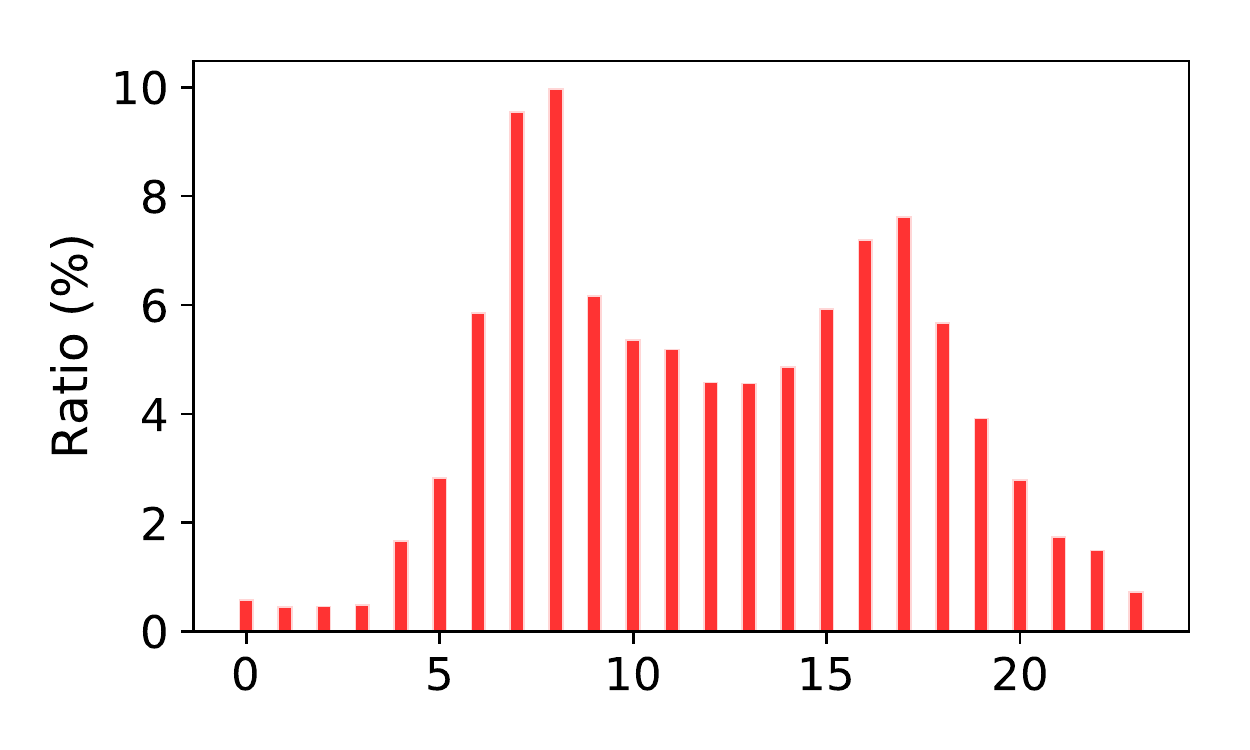}
        (b) Hour of Day (weekdays)
    \endminipage\hfill
    \minipage{0.32\textwidth}
        \centering
        \includegraphics[width=\linewidth]{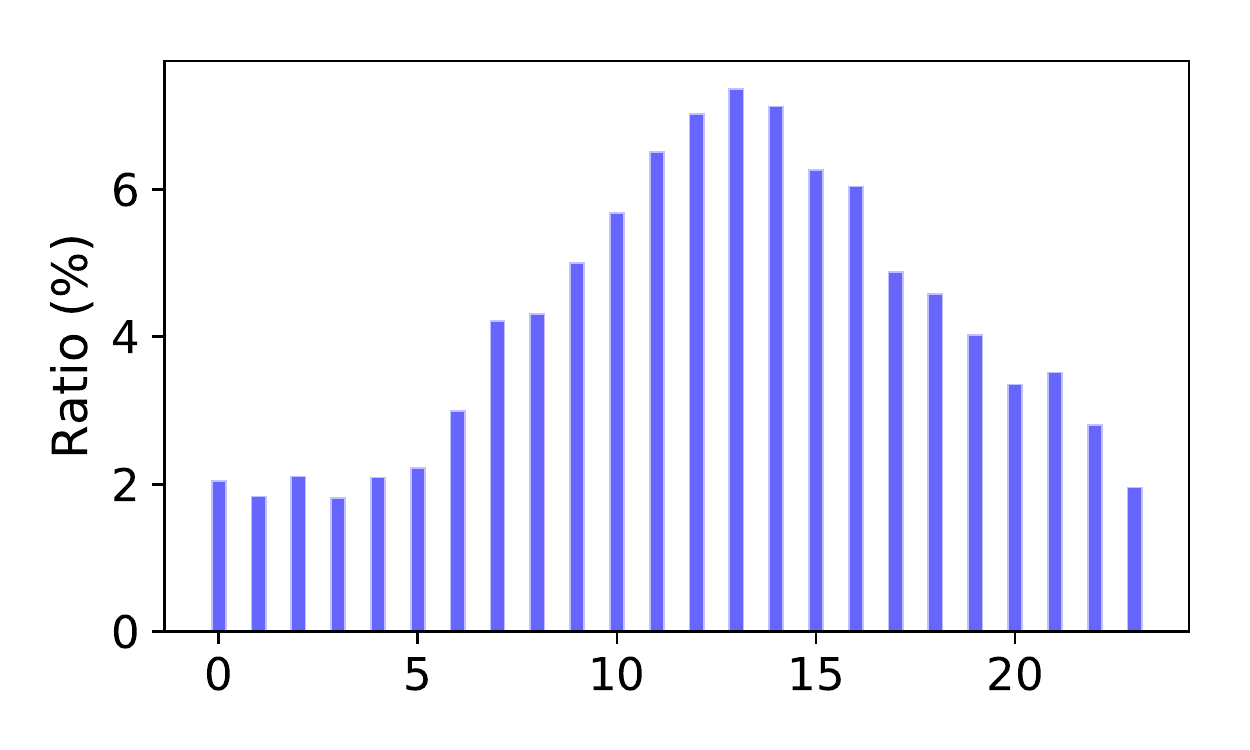}
        (c) Hour of Day (weekends)
    \endminipage\hfill
    \minipage{0.32\textwidth}
        \centering
        \includegraphics[width=\linewidth]{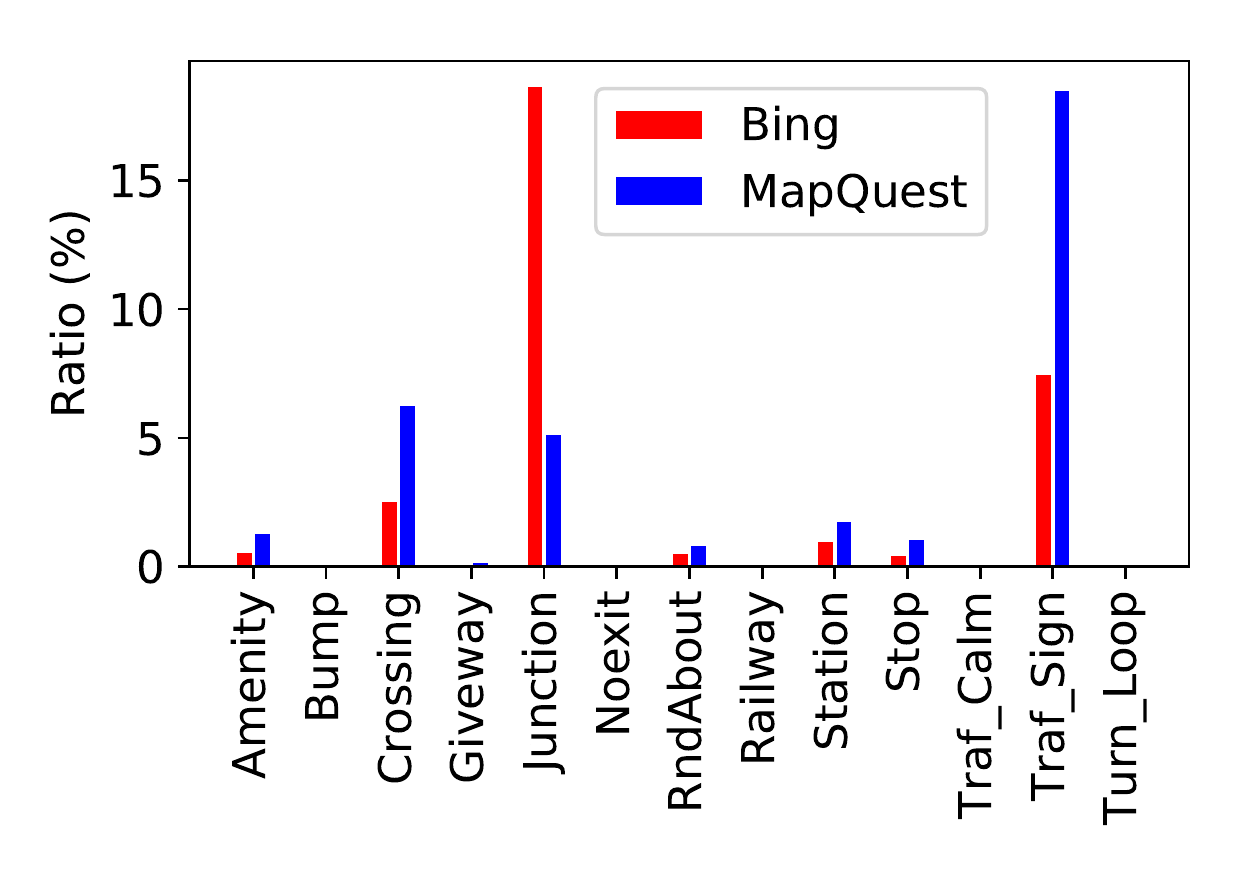}
        (d) points-of-interest Distribution
    \endminipage\hfill
    \minipage{0.32\textwidth}
        \centering
        \includegraphics[width=\linewidth]{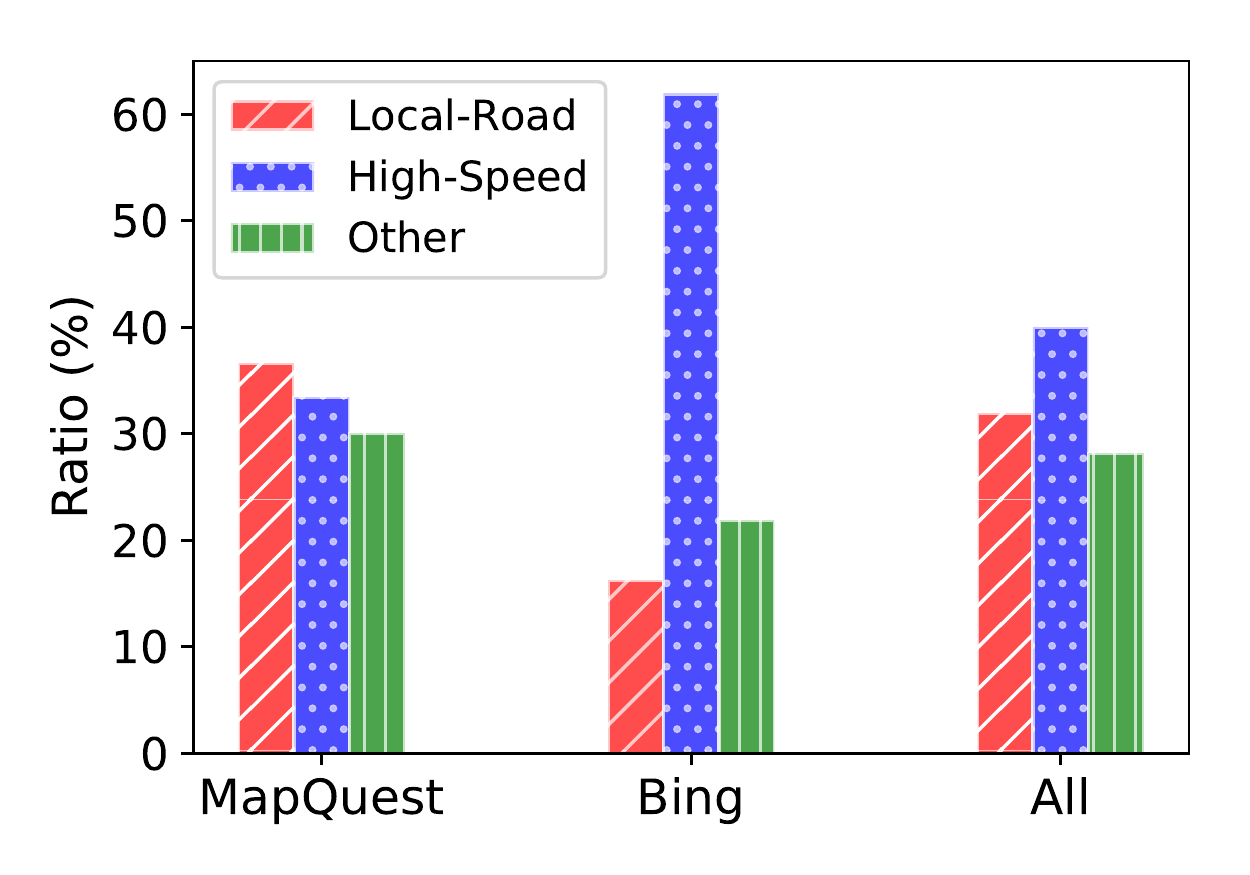}
        (e) Road-type Distribution
    \endminipage\hfill
    \minipage{0.32\textwidth}
        \centering
        \includegraphics[width=\linewidth]{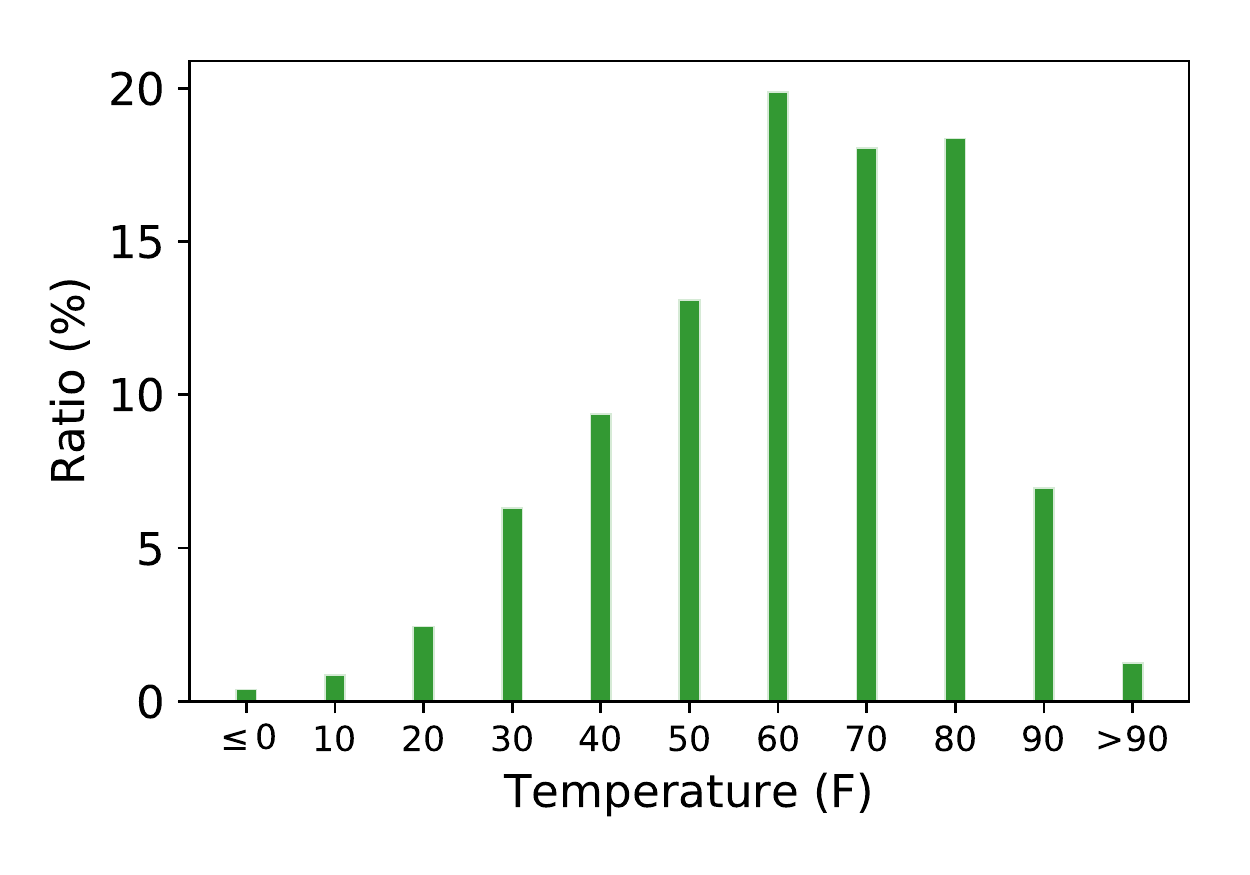}
        (f) Temperature Distribution
    \endminipage\hfill
    \minipage{\textwidth}
        \centering
        \caption{Characteristics of US-Accidents dataset, in terms of time analysis (a)--(c), points-of-interest-based augmentation distribution analysis (d), map-matching-based road type coverage analysis (e), and temperature analysis (f).}
        \label{fig:all_profilings}
    \endminipage\hfill
\end{figure*}

Using the process described in Section~\ref{sec:data_process}, we created a countrywide dataset of traffic accidents, which we name {\em US-Accidents}. US-Accident contains about $2.25$ million cases of traffic accidents that took place within the contiguous United States from February 2016 to March 2019. 
Table~\ref{tab:data_facts} shows the important details of US-Accidents. Also, Figure~\ref{fig:all_profilings} provides more details on characteristics of the dataset. Figure~\ref{fig:all_profilings}-(a) shows the daily distribution of traffic accidents, where significantly more accidents were observed during the weekdays. Based on parts (b) and (c) of Figure~\ref{fig:all_profilings}, it can be observed that the hourly distribution during weekdays has two peaks (8am and 5pm), while the weekend distribution shows a single peak (1pm). Figure~\ref{fig:all_profilings}-(d) demonstrates that most of the accidents took place near junctions or intersections (crossing, traffic signal, and stop). MapQuest tends to report more accidents near intersections, while Bing reported more cases near junctions. This shows the complementary behavior of these APIs and the comprehensiveness of our dataset. 
Figure~\ref{fig:all_profilings}-(e) describes distribution of road types, extracted from the map-matching results (i.e., street names). Here we note that about $32\%$ of accidents happened on or near local roads (e.g., streets, avenues, and boulevards), and about $40\%$ took place on or near high-speed roads (e.g., highways, interstates, and state roads). We also note that Bing reported more cases on high-speed roads. Finally, the period of day data shows that about $73\%$ of accidents happened after sunrise (or during the day). 

\begin{table}[ht]
    \small
    \setlength\tabcolsep{1pt}
    \centering
    \caption{\small US-Accidents: details as of March 2019.} 
    \begin{tabular}{|>{\columncolor[gray]{0.9}}c|c|}
        \hline
        Total Attributes &  45\\
        \hline
        Traffic Attributes (10) &  \begin{tabular}{@{}c@{}} id, source, TMC \cite{tmc}, severity, start\_time, end\_time, \\ start\_point, end\_point, distance, and description \end{tabular} \\
        \hline
        Address Attributes (8) & \begin{tabular}{@{}c@{}} number, street, side (left/right), city, \\ county, state, zip-code, country  \end{tabular}\\
        \hline
        Weather Attributes (10) & \begin{tabular}{@{}c@{}} time, temperature, wind\_chill, humidity,\\ pressure, visibility, wind\_direction, wind\_speed,\\precipitation, and condition (e.g., rain, snow, etc.) \end{tabular}\\
        \hline
        POI Attributes (13) & All cases in Table~\ref{tab:poi_types}\\
        \hline
        Period-of-Day (4) & \begin{tabular}{@{}c@{}} Sunrise/Sunset, Civil Twilight, \\Nautical Twilight, and Astronomical Twilight\end{tabular}\\
        \hline
        \hline
        Total  Accidents &  2,243,939\\
        \hline
        \# MapQuest Accidents & 1,702,565 (75.9\%) \\
        \hline
        \# Bing Accidents & 516,762 (23\%)\\
        \hline
        \# Reported by Both & 24,612 (1.1\%) \\
        \hline
        Top States & \begin{tabular}{@{}c@{}}  California (485K), Texas (238K), Florida (177K), \\ North Carolina (109K), New York (106K) \end{tabular}\\
        \hline
    \end{tabular}
    \label{tab:data_facts}
    
\end{table}

To further compare the US-Accidents dataset with the other publicly available sources, Table~\ref{tab:compare_datasets} provides some details in this regard. Regarding the size of data, US-Accidents is by far the largest available set. UK Accidents \cite{uk_accidents} is the only publicly available countrywide dataset, and its yearly reports are of about $100 K$ accidents \footnote{Based on \cite{uk_accidents}, there is no data reported for 2008.}. US-Accidents, however, contains about $750 K$ accidents for each year. US-Accidents also provides many more details for each accident record than (say) New York Accidents \cite{ny_accidents}. 

\begin{table}[ht]
    \setlength\tabcolsep{1pt}
    \small
    \centering
    \caption{\small Comparing publicly available, large-scale accident datasets}
    \begin{tabular}{c|c|c|c|c|c}
        \hline
        \rowcolor{Gray2}
        \textbf{Dataset} & \textbf{State} & \textbf{Country} & \textbf{Time} & \textbf{Size} & \textbf{Source}\\
        \hline
         UK Accidents \cite{uk_accidents}              &  -- & UK & 2000--2016 & 1.6 M & Police Reported\\
         Seattle Crash Report \cite{seattle_accidents} &  WA & -- & 2004--2018 & 208 K & Police Reported\\
         Iowa Accidents \cite{ia_accidents}            &  IA & -- & 2008--2018 & 557 K & Iowa DOT\\
         New York Accidents \cite{ny_accidents}        &  NY & -- & 2014--2016 & 1.65 M & NYS DMV\\
         Maryland Accidents \cite{md_accidents}        &  MD & -- & 2015--2018 & 400 K & Police Reported\\
        \rowcolor{LightCyan}
         US-Accidents                                  &  -- & US & 2016--2019 & 2.25 M & Streaming Data\\
         \hline
    \end{tabular}
    \label{tab:compare_datasets}
\end{table}

\subsection{Applications of the Dataset}
US-Accidents may be used for applications such as real-time accident prediction; studying accident hotspot locations; casualty analysis (extracting cause and effect rules to predict accidents); or studying the impact of precipitation or other environmental stimuli on accident occurrence. Given the scale of data, researchers may utilize this dataset to derive a variety of insights which can benefit applications such urban planning and improving transportation infrastructures. In our own recent study, we employed the US-Accidents dataset along with the other traffic and weather events to perform pattern discovery over large-scale geo-spatiotemporal data, and revealed a variety of insights in terms of {\em propagation} and {\em influential} patterns \cite{moosavi2019short}. 
\section{Conclusion and Future Work}
\label{sec:conclusion}
This paper describes US-Accidents, a unique, publicly available motor vehicle accident dataset, and its process of creation -- that includes several important steps such as real-time traffic data collection, data integration, and multistage data augmentations using map-matching, weather, period-of-day, and points-of-interest data. To the best of our knowledge, US-Accidents is the first countrywide dataset of this scale, containing about $2.25$ million traffic accident records collected for the contiguous United States over three years. From this dataset, we were able to derive a variety of insights with respect to the location, time, weather, and points-of-interest of an accident. We believe that US-Accidents provides a context for future research on traffic accident analysis and prediction. In terms of our own future work, we plan to employ this dataset to perform real-time traffic accident prediction. 

\section*{Acknowledgment}
This work is supported by a grant from the Ohio Supercomputer Center (PAS0536). 

\bibliographystyle{ACM-Reference-Format}
\bibliography{main.bbl}

\end{document}